\documentclass{article} 
\usepackage{epsfig} 
\usepackage{cite}  
\def\e{\epsilon}
\def\d{\hbox{d}}

\def\nn{\nonumber}

\begin{document} 
\unitlength1cm 
\begin{titlepage} 
\vspace*{-1cm} 
\begin{flushright} 
ZU--TH 19/03\\
IPPP/03/75\\
DCPT/03/150\\
hep-ph/0311276\\
November 2003
\end{flushright} 
\vskip 3.5cm 

\begin{center} 
{\Large\bf Four-Particle Phase Space Integrals in
Massless QCD}
\vskip 1.cm 
{\large  A.~Gehrmann--De Ridder}$^{a}$, {\large  T.~Gehrmann}$^a$ 
and {\large G.~Heinrich}$^b$ 
\vskip .7cm 
{\it $^a$ Institut f\"ur Theoretische Physik, Universit\"at Z\"urich,
Winterthurerstrasse 190, CH-8057 Z\"urich, Switzerland} 
\vskip .4cm 
{\it $^b$ Institute for Particle Physics Phenomenology, University of Durham,
South Road,\\ Durham DH1 3LE, England} 
\end{center} 
\vskip 2.6cm 

\begin{abstract} 
The inclusive four-particle phase space integral of any 
$1\to 4$ matrix element in massless QCD contains 
divergences due to the soft and collinear emission of up to two particles in
the final state. We show that any term appearing in this phase space 
integral can be expressed as linear combination of only four master integrals.
These four master integrals are all computed in dimensional regularisation
up to their fourth order
terms, relevant to next-to-next-to-leading order jet calculations, both 
in an analytic form and purely numerically. New analytical and numerical 
techniques are developed in this context. 
We introduce the tripole parametrisation 
of the four-parton phase space. Furthermore, we exploit 
 unitarity relations between multi-parton phase space 
integrals and multi-loop integrals. For the numerical calculation, the 
iterated sector decomposition of loop integrals is extended to phase space 
integrals. The results in this paper lead to infrared subtraction terms 
needed for the double real radiation contributions to jet physics 
in $e^+e^-$ annihilation at the 
next-to-next-to-leading order. 
\end{abstract} 
\vfill 
\end{titlepage} 
\newpage 

\renewcommand{\theequation}{\mbox{\arabic{section}.\arabic{equation}}} 

\section{Introduction}
\setcounter{equation}{0}

In recent years, data on jet observables from 
LEP, HERA and the Tevatron have reached an impressive level of 
experimental precision. Using these data for a determination of 
the strong coupling constant, or as a measurement of parton distributions 
in the colliding hadrons, it turns out that the error on these extractions is 
largely dominated by the uncertainty on the underlying theoretical 
calculations, which are accurate only to the next-to-leading order (NLO) 
in perturbative QCD. 

In order to improve on this situation, 
theoretical calculations of jet observables accurate to the 
next-to-next-to-leading order (NNLO) are mandatory. For each observable,
they require several ingredients:
To compute the corrections to an $n$-jet
observable, one needs the two-loop $n$-parton matrix elements, 
the one-loop ($n$+1)-parton matrix elements and the tree level ($n$+2)-parton 
matrix elements. In the recent past, enormous progress has been made 
especially on the calculation of two-loop $2\to 2$ and $1\to 3$ 
QCD matrix elements, which are now known for all 
massless parton--parton scattering 
processes~\cite{m1,m2,m3,m4} relevant to hadron colliders  as well as for 
$\gamma^*\to q\bar q g$~\cite{3jme,muw2} and its crossings~\cite{ancont}. 
Most of these results 
were even verified independently by different groups.
For the corresponding partonic processes, 
the one-loop matrix elements with one 
additional parton and the tree-level matrix elements with two more partons 
are also known and form part of NLO programs for 
$1\to 4$~\cite{nplusone1} and $2\to 3$ reactions~\cite{nplusone2}. 
Since these matrix elements lead to infrared
singularities due to one or two partons becoming theoretically
unresolved (soft or collinear), one needs to find one- and
two-particle subtraction terms, which account for these singularities 
in the matrix elements, and are sufficiently simple to be integrated
analytically over the unresolved phase space. One-particle
subtraction at tree level is well understood from NLO 
calculations\cite{eks,gg,cs} and general algorithms are available for 
one-particle subtraction at one loop\cite{onel,kos1,wz1}, in a 
form that could recently be integrated 
analytically~\cite{kos1,wz1}. Tree-level
two-particle subtraction terms have been extensively studied in the 
literature\cite{twot,kosower,wz2},
their integration over the unresolved phase space 
 could however up to now  be performed 
only in one particular infrared subtraction scheme 
(the hybrid subtraction method) in the
calculation of higher order corrections to the photon-plus-one-jet
rate in $e^+e^-$ annihilation~\cite{ggam1,ggam}. The same techniques
(and the same scheme) were used 
very recently in the rederivation of the time-like gluon-to-gluon
splitting function from splitting 
amplitudes~\cite{dak}.
A general two-particle subtraction
procedure, consisting of subtraction terms which can be safely implemented 
numerically and at the same time are sufficiently simple 
to be integrated analytically  is still lacking at the moment. It is 
the aim of the present paper to contribute to such a method by 
%systematizing 
classifying and computing all 
four-particle phase space integrals 
relevant to $1\to 4$ parton reactions in massless QCD. 
  
One of the most widely used infrared subtraction schemes at NLO is 
the dipole formalism established by Catani and Seymour~\cite{cs}.
The formalism relies on an exact factorisation of the $n$-particle 
phase space into an ($n$--1)-particle phase space and a so-called dipole 
phase space, which is, up to a normalisation factor, equal to a
massless three-particle phase space. In this formalism, one derives 
(process-independent) dipole subtraction terms for all single
unresolved partonic configurations, which are
then integrated over the dipole phase space (corresponding to a three-parton 
configuration of emitter, unresolved parton and spectator). If combined 
appropriately, these integrals of the subtraction terms over the dipole 
phase space can be related to integrals of $1\to 3$ tree level QCD matrix 
elements integrated over the three-particle phase space. 

At NNLO, one 
encounters configurations of two unresolved partons between emitter and 
spectator. As will be shown below in Section~\ref{sec:tripole}, 
these can be accommodated into a tripole formulation, relying on the 
exact factorisation of the 
 $n$-particle 
phase space into an ($n$--2)-particle phase space and a tripole  phase space. 
This tripole phase space is proportional to the four-particle phase space. 
Based on this observation, one might envisage that all double unresolved 
partonic configurations which correspond to a single 
emitter-spectator pair may be expressed by appropriate tripole subtraction 
terms. To facilitate their analytic integration over the tripole phase space,
these may be further combined to relate to integrals of $1\to 4$ 
tree-level QCD matrix 
elements integrated over the four-particle phase space. In this paper, we 
show that all integrals of this type can be expressed as linear combinations 
of only four so-called master integrals, which we compute both analytically 
and numerically. 

The reduction to master integrals follows largely methods developed for 
multi-loop integrals, and is briefly presented in Section~\ref{sec:red}. 
To compute the resulting master integrals analytically in 
Section~\ref{sec:an}, we 
develop first the tripole formulation of the four-particle phase space, 
which is then used to calculate two of the master integrals. 
We then show that the remaining two master integrals 
can be inferred from unitarity relations 
between known multi-loop integrals and phase space integrals. 
Further, one of the master integrals calculated explicitly,  
as well as a reducible integral, are rederived from unitarity relations
as crosschecks. 

In Section~\ref{sec:num}, the master integrals are calculated numerically. 
The infrared poles are isolated by an automated sector decomposition 
algorithm which already has 
proven useful in the calculation of multi-loop 
integrals~\cite{Binoth:2000ps,Binoth:2003ak}, and which has been 
extended in~\cite{Heinrich:2002rc} and in the present work 
to tackle infrared divergent integrals over a multi-parton  
phase space as well.

A summary and conclusions are given in Section~\ref{sec:conc}.
The Appendix contains formulae relevant to the massless 
($1\to n$)-parton phase space.

\section{Kinematics and Notation}
\setcounter{equation}{0}
\label{sec:kin}

In this paper, we consider the decay of a massive particle $Q$
($Z$-boson, off-shell photon or Higgs-boson) 
into four massless QCD partons ($P_1,\ldots 
P_4$: $q\bar q gg$, $q\bar q q\bar q$ or $gggg$):
\begin{equation}
Q (q) \to P_1(p_1) + P_2(p_2) + P_3(p_3) + P_4(p_4) \; ,
\end{equation}
where $q, p_1,\ldots,p_4$ denote the momenta of the particles.
The kinematics of this process are uniquely defined by the six Lorentz 
invariants $s_{ij} = 2 p_i \cdot p_j$ formed by any pairs of partons. These 
appear in the numerator and denominator (propagators) of QCD matrix elements.  
Only five of them are independent, the sixth being fixed by energy-momentum 
conservation
$$ q^2 = s_{12} + s_{13} + s_{14} + s_{23} + s_{24} + s_{34} \;.
$$
Besides these, QCD matrix elements also contain 
propagators yielding triple invariants 
$$
s_{ijk} = s_{ij} + s_{ik} + s_{jk} \; .
$$
Without loss of generality, we can always choose $P_1$ and $P_2$ such that 
the associated scalar product $s_{12}$ and triple invariants 
$s_{123}$, $s_{124}$ do not appear as a propagator in a given term of 
the squared matrix element.  

Any single term in a squared four-particle QCD matrix element can then be 
expressed as 
\begin{equation}
T(S_1^{n_1} \ldots S_q^{n_q}; D_1^{m_1} \ldots D_t^{m_t})
 = \frac{S_1^{n_1} \ldots S_q^{n_q}}{D_5^{m_1} \ldots D_t^{m_t}}\,
\label{eq:singleterm}
\end{equation}
where the $S_l$ are the Lorentz invariants $s_{ij}$ and the $D_l$ are 
the propagators, being either invariants $s_{ij}$ or triple invariants
$s_{ijk}$. The $n_i$ and $m_i$ are positive integers.
It is assumed that whenever possible, 
scalar products in the numerator are expressed as 
linear combinations of the propagators, such that only a minimal number of
different scalar products is present in the numerator. In fact, one finds 
that for $t-4$ different propagators, only $9-t$ different irreducible 
scalar products exist.
The integral of the term (\ref{eq:singleterm}) over the 
four-particle phase space is denoted by 
\begin{equation}
I_T(S_1^{n_1} \ldots S_q^{n_q}; D_5^{m_5} \ldots D_t^{m_t}) = 
\int \d PS_4 \; T(S_1^{n_1} \ldots S_q^{n_q}; D_5^{m_5} \ldots D_t^{m_t}) \;.
\label{eq:itmaster}
\end{equation}
The phase space measure $\d PS_4$ is defined in (\ref{eq:r4}) 
in  the appendix. 

The topology of $I_T$ (interconnection of propagators and external momenta) 
is uniquely determined by the set of propagators $(D_5,\ldots ,D_t)$. In 
order to visualise the topology of $I_T$ in the form of a cut diagram, we 
use the 
Cutkosky rules\cite{cutcu,babis} to introduce the four cut-propagators
\begin{equation} 
\frac{1}{D_i} 
= 2\pi i \delta^+(p_i^2) = \frac{1}{p_i^2 + i0} - \frac{1}{p_i^2-i0}\;  
\qquad (i= 1,\ldots 4).
\end{equation}
With the help of the cut-propagators, the four-particle phase space 
measure in 
(\ref{eq:itmaster}) can be written as 
\begin{equation}
\d PS_4 =  i^{-4} \frac{\d^d p_1}{(2\pi)^d} \,
\frac{\d^d p_2}{(2\pi)^d} \,
\frac{\d^d p_3}{(2\pi)^d} \,
\frac{\d^d p_4}{(2\pi)^d} \,(2\pi)^d \,\delta^d (q-p_1-p_2-p_3-p_4) \; 
\frac{1}{D_1\, D_2\, D_3\, D_4}\; ,
\end{equation}
such that the integral $I_T$ can be expressed as   
a cut through a three-loop vacuum polarisation diagram. 
Any integral $I_T$ for a given topology (fixed set of $t-4$ propagators) 
can be attributed to a class of integrals $I_{T(t,r,s)}$ of identical 
numerator mass dimension 
$s=\sum_{i=1}^{9-t} n_i$ and denominator mass dimension 
$r=\sum_{i=1}^t m_i$. It should be noted that we also allow the cut-propagators
to appear in general with integer powers larger than unity, without further 
specifying the physical meaning of those. 
From combinatorics, one finds that the class $I_{T(t,r,s)}$ contains 
$$
N_{t,r,s} = { r-1 \choose r-t } \, { 8-t+s \choose s}
$$
different integrals.

\section{Reduction to Master Integrals}
\setcounter{equation}{0}
\label{sec:red}

The number $N(I_{t,r,s})$ of the integrals grows quickly as $r, s$ 
increase, but the integrals are not all linearly independent. 
These linear relations are the so-called integration-by-part (IBP) 
identities, which  were originally 
formulated for multi-loop two-point functions~\cite{chet1,chet2} with ordinary
loop propagators, but can be safely extended to multi-loop integrals with 
cut propagators~\cite{babis}. 
These identities follow from the fact that the integral over the 
total derivative with respect to any loop momentum vanishes in
dimensional regularisation
\begin{equation}
\int \frac{\d^d k}{(2\pi)^d} \frac{\partial}{\partial k^{\mu}}
J(k,\ldots)  = 0,
\end{equation} 
where $J$ is any combination of propagators, scalar products
and loop momentum vectors. $J$ can be a vector or tensor of any rank. 
The IBP identities for three-loop two-point functions were derived and 
solved symbolically
a long time ago~\cite{chet2} in the context of the calculation of the
three-loop $\beta$-function. The solution of these identities was implemented 
into the program package {\tt Mincer}~\cite{mincer} written in the algebraic
programming language {\tt Form}~\cite{form}.
{\tt Mincer}  was widely used 
in recent years to compute a large number of multi-loop corrections 
which correspond to three-loop two-point functions or can be related to 
them by expansion. Reviews of these results can be found 
in~\cite{harlrev,steinhrev}. 

In recent times, IBP-type relations between multi-loop integrals 
with more than two legs have been studied extensively. In particular, it 
was found that scalar
multi-loop integrals with three or more legs fulfil, besides 
the ordinary IBP relations,
also relations following from the invariance of the 
integrals under an infinitesimal Lorentz transformation. These Lorentz 
invariance (LI) identities, combined with the IBP identities, were used 
extensively in the computation of  two-loop four-point 
matrix elements~\cite{m1,m2,m3,m4,3jme,muw2}.
In a related development, it was observed that 
the IBP identities  for Feynman integrals with the same total number of 
external and loop momenta are equivalent~\cite{baikov}, a development that 
was used soon thereafter to compute two-loop three-point 
functions~\cite{harlanderhiggs}. 

Exploiting the observation that differentiations involved in the 
integration-by-parts identities are insensitive to 
the imaginary parts of the two 
terms in a cut-propagator~\cite{babis}, one can generalise the IBP method 
to phase space integrals over tree-level or loop matrix elements. 
In the solution of the IBP equations for a given cut integral, one obtains 
integrals with one or more of the cut propagators eliminated. These integrals 
can be discarded, since they do not contribute anymore to the same cut of 
the underlying multi-loop diagram. 

To derive the reduction of any given integral $I_T$ of the form 
(\ref{eq:itmaster}), we use the same strategy~\cite{laporta}
 as applied in the reduction of 
the two-loop four-point functions in~\cite{gr} by generating all IBP identities
for integrals up to the maximum values of $r$ and $s$ required for 
the integration of QCD four-particle matrix elements. From dimensional 
counting, one finds $r\leq 8$, $s\leq 3$. The solution algorithm~\cite{laporta}
 proceeds by solving all identities up to these values using the computer 
algebra languages {\tt Form}~\cite{form} and {\tt Maple}~\cite{maple}.

After carrying out the reduction for all topologies that can be formed 
from the invariants and triple invariants involving  the 
four final state momenta, we find that all inclusive 
four-particle phase space integrals of massless QCD matrix elements can be 
expressed as a linear combination of four master integrals. 
These four master integrals are:
\begin{eqnarray}
R_4 & \equiv & \parbox{1.4cm}{\epsfig{file=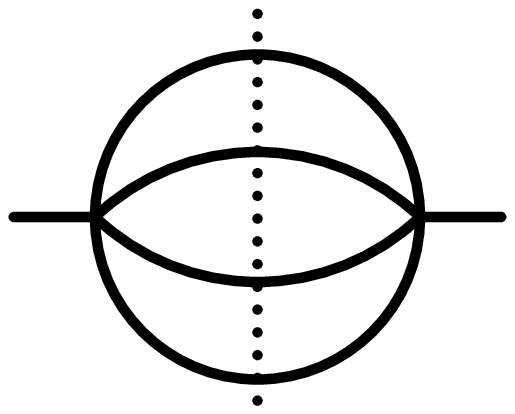,width=1.4cm}}= 
\int \d PS_4 = P_4 \\
R_6 & \equiv & \parbox{1.4cm}{\epsfig{file=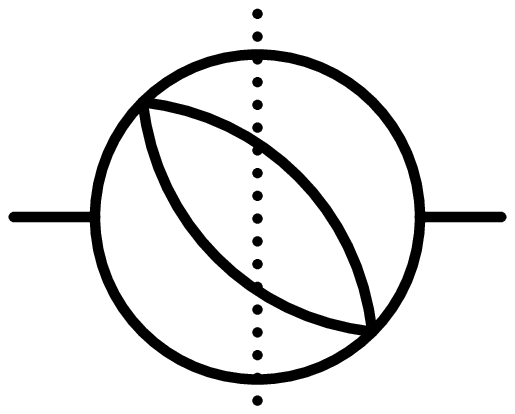,width=1.4cm}}=
\int \d PS_4 \frac{1}{s_{134}s_{234}}\\
R_{8,a} & \equiv & \parbox{1.4cm}{\epsfig{file=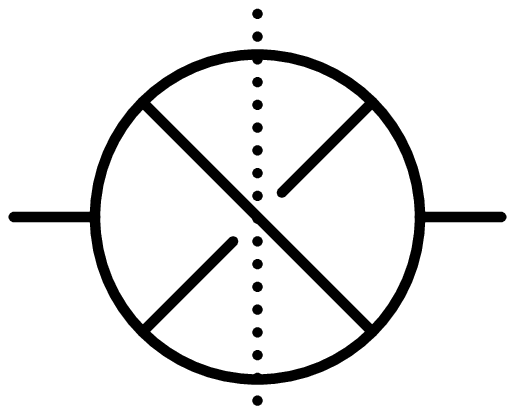,width=1.4cm}}=
\int \d PS_4 \frac{1}{s_{13}s_{23}s_{14}s_{24}}\\
R_{8,b} & \equiv & \parbox{1.4cm}{\epsfig{file=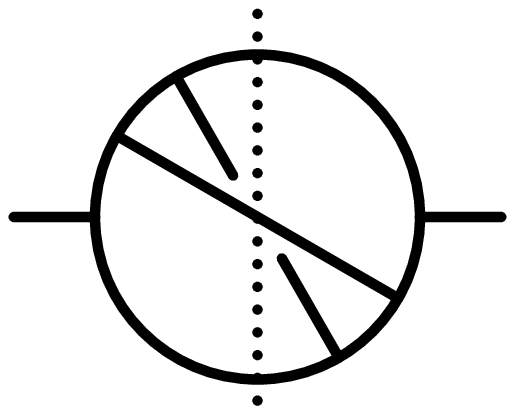,width=1.4cm}}=
\int \d PS_4 \frac{1}{s_{13}s_{134}s_{23}s_{234}}\;.
\end{eqnarray}
The reduction procedure generates coefficients containing 
explicit poles of up to $1/\e^4$ in front of $R_4$ and up to $1/\e^2$ in front of 
$R_6$, while the coefficients multiplying $R_{8,a}$ and $R_{8,b}$ are finite. 

In the following two sections, we will compute these four master integrals 
both analytically and numerically to a sufficiently high order in $\e$ 
required for the calculation of integrated four-particle QCD matrix elements.

\section{Analytic Calculation of Master Integrals} 
\setcounter{equation}{0} 
\label{sec:an} 

\subsection{Tripole Phase Space}
\label{sec:tripole}
\newcounter{myr4}
\newcounter{myr6}
\newcounter{myr8a}
\newcounter{myr8b}

The dipole formalism~\cite{cs}
 of Catani and Seymour is a subtraction formalism for the 
calculation of jet cross sections at NLO.  
The main features of this formalism are factorisation formulae for 
matrix elements and phase space which locally reproduce single soft 
and single collinear behaviour of the cross sections in the appropriate
limits. The corresponding dipole contributions to the cross sections can be
integrated analytically and numerically over the whole phase space. 
In particular, the exact factorisation of the three-parton phase space into 
a dipole phase space and a two-parton phase space is obtained by 
redefining a set of three massless on-shell momenta (emitter, unresolved
 parton, spectator) into two on-shell massless momenta. This procedure 
was further generalised by Kosower~\cite{kosower}, also accounting 
for the recombination of more than three partonic momenta. 
In~\cite{kosower} a non-linear remapping of momenta, which smoothly 
interpolates between all potentially singular regions (as required for 
the numerical implementation) was introduced. Such a non-linear remapping 
is however inappropriate for analytic integration, and we shall derive a 
linear remapping below. 

As an extension of~\cite{cs},
we introduce a parametrisation of momenta allowing  
the factorisation 
of a massless $n$-parton phase space into a massless ($n$--2)-parton phase 
space times a tripole phase space factor:
\begin{equation}
\d PS_n(p_1,\ldots,p_n) = \d PS_{n-2}(p_1,\ldots,p_{n-4},
\tilde{p}_{n-3},\tilde{p}_{n-2})
 \d PS_T \; ,
\end{equation}

In particular,
it enables to factorise exactly 
the phase space of four on-shell parton momenta 
($p_{1}..p_{4}$),
which can also denote the four-particle contribution to an 
($n>4$)-particle phase space, into a two-parton 
phase space times a tripole phase space factor 
\begin{equation}
\d PS_{4}(p_1,\ldots,p_4) = 
\d PS_{2}(p_{\widetilde{134}},p_{\tilde{2}})
\; \d PS_T. 
\label{eq:tripole1}
\end{equation}
In this parametrisation, $p_1$ is the momentum of the emitter, $p_2$ the one of
the spectator, while $p_3$ and $p_4$ are the momenta of the unresolved 
partons.

The tripole momenta ($p_{\widetilde{134}},p_{\tilde{2}}$ ) 
are combinations of four on-shell parton momenta. They are defined by
\begin{eqnarray}
p_{\widetilde{134}}^{\mu}&=&p_{1}^{\mu} +p_{3}^{\mu} +p_{4}^{\mu} 
- \frac{y_{134,2}}{1-y_{134,2}} p_{2}^{\mu} \;, \nonumber \\
p_{\tilde{2}}^{\mu}&=&\frac{1}{1-y_{134,2}} p_{2}^{\mu} \;,
\end{eqnarray}
where the dimensionless variable $y_{134,2}$ is 
\begin{equation}
y_{134,2}=\frac{p_1.p_3 +p_1.p_4
+p_3.p_4}{p_1.p_3+p_1.p_4+p_3.p_4+p_1.p_2 
+p_2.p_3 +p_2.p_4}.
\end{equation}
The tripole momenta are on-shell ($p_{\widetilde{134}}^{2}=p_{\tilde{2}}^{2}=0$ ) 
and momentum conservation is implemented exactly
$$
p_{1}^{\mu}+p_{2}^{\mu}+p_{3}^{\mu}+p_{4}^{\mu}=p_{\widetilde{134}}^{\mu} +
p_{\tilde{2}}^{\mu}.
$$

To derive the tripole factorisation of $\d PS_4$, we start from its 
definition in terms of the invariants $s_{ij}$
\begin{eqnarray*}
\d PS_{4} & = & (2\pi)^{4-3d}\;(q^2)^{3-\frac{d}{2}}\;2^{1-2d}
(-\Delta_{4})^{\frac{d-5}{2}}\;\Theta(-\Delta_{4})\;
\;\delta(s_{12}+ s_{13}+s_{14}+s_{23}+s_{24}+s_{34}-q^{2})
 \\
& & \int {\rm d}\Omega_{d-1}\;{\rm d}\Omega_{d-2}\;{\rm d}\Omega_{d-3}
{\rm d}s_{12}\;{\rm d}s_{13}\;{\rm d}s_{14}\;{\rm d}s_{23}\;{\rm d}s_{24}\,
{\rm d}s_{34}\;,
\end{eqnarray*}
where the four-particle Gram determinant $\Delta_{4}$ is given by
\begin{eqnarray}
+\Delta_{4} & = & 
\bigg[{s_{12}}^{2}{s_{34}}^{2}\; +\;
{s_{13}}^{2}{s_{24}}^{2}\;+\;{s_{14}}^{2}{s_{23}}^{2}
\nonumber\\
& & -2\bigg(s_{12}s_{23}s_{34}s_{14}\;+\;s_{13}s_{23}s_{24}s_{14}\;
+\;s_{12}s_{24}s_{34}s_{13} \bigg)\bigg]
\nn \\
&=& \lambda \left(s_{12}s_{34},s_{13}s_{24},s_{14}s_{23}\right)\label{eq:delta4}
\end{eqnarray}
and $\lambda$ is the K\"allen function, 
$\lambda(x,y,z)=x^2+y^2+z^2-2(xy+xz+yz)$. 

In order to obtain the massless four-parton phase space $ \d PS_{4}$ 
in terms of dimensionless invariants defined with 
$p_{\widetilde{134}},p_{\tilde{2}}$ as given in (\ref{eq:tripole1}), 
the invariants $s_{ij}$  need to be substituted as follows
\begin{eqnarray}
s_{23}&=&(1-y_{134,2})s_{\tilde{2}3}=(1-y_{134})\,q^2\,z_1\;, \nn\\
s_{24}&=&(1-y_{134,2})s_{\tilde{2}4}=(1-y_{134})\,q^2\,z_2\;, \nn\\
s_{12}&=&(1-y_{134,2})s_{1\tilde{2}}=(1-y_{134})\,q^2\,(1-z_1-z_2)\;.
\end{eqnarray} 
The last simplification occurs since,
in the case of only four partons in the final state,
one has
\begin{equation}
2\,p_{\widetilde{134}}.p_{\tilde{2}}= \;s_{\widetilde{1234}}=s_{1234}=q^2
\label{eq:mom2}
\end{equation} 
and $y_{134,2}$ is the dimensionless variable $y_{134}={s_{134}}/{q^2}$.
The fractions of momenta $z_{i}$ are defined by:
\begin{eqnarray}
z_{1}&=&\frac{p_{3}.p_{\tilde{2}}} {p_{\widetilde{134}}.p_{\tilde{2}}} \;,\nn\\
z_{2}&=&\frac{p_{4}.p_{\tilde{2}}}{p_{\widetilde{134}}.p_{\tilde{2}}}\;.
\end{eqnarray}
In the tripole phase space $\d PS_{T}$, 
the variables $s_{13},s_{14}$ and $s_{134}$ remain unchanged and 
the triple invariant $s_{134}$ is used  instead of $s_{34}$.

From the  four-parton phase space, one can easily factorise a two-parton
phase space factor $\d PS_2$ as defined in the Appendix
using (\ref{eq:mom2}). The integration variables in
$\d PS_4$ become
\begin{displaymath}
{\rm d} s_{12}\, {\rm d}s_{13}\,
 {\rm d}s_{14}\,{\rm d}s_{23}\,{\rm d}s_{24}\,
{\rm d}s_{134}
=(1-y_{134})^2\;(q^2)^5 \,{\rm d}{s}_{1234} \,{\rm d}y_{13} \,{\rm d}y_{14}\,
{\rm d}z_{1} \, {\rm d}z_{2}\,{\rm d}y_{134}\;,
\end{displaymath}
where the ${\rm d}s_{1234}$ 
on the right-hand side becomes part of the two-parton phase space 
while the other integration variables contribute to the tripole phase
space factor $\d PS_T$. 
Furthermore, 
the Gram determinant $\Delta_{4}$ is  rescaled as 
$$
\Delta_{4}=(1-y_{134})^2\;(q^2)^4 \; \Delta'_4\;,
$$
with 
\begin{equation}
\Delta'_4 = 
\lambda
\left(y_{12}(y_{134}-y_{13}-y_{14}),y_{13}z_2,y_{14}z_1\right)\; .
\end{equation}
The factorised form of the four-parton phase space yielding the
tripole phase space factor then reads:
\begin{equation}
\d PS_{4}=  \d PS_{2} \; \d PS_{T}\;,
\label{eq:fac} 
\end{equation}
with
\begin{eqnarray}
\d PS_{T} =  (q^2)^{2-2\e} 
\frac{2^{-10}\;\pi^{-5+2\e}}{\Gamma(1-2\e)} \;
 (1-y_{134})^{1-2\e}(-\Delta'_{4})^{-1/2-\e}\,\Theta(-\Delta'_{4})\,
{\rm d}y_{13}\; {\rm d}y_{14}\;
{\rm d}z_{1} \; {\rm d}z_{2}\;{\rm d}y_{134}\;.
\end{eqnarray}

This tripole phase space factor $\d PS_{T}$ is proportional to the 
four-parton phase space, since the two-parton phase space is a constant. 
It is also  proportional to the triple collinear phase
space factor given in \cite{ggam} obtained by requiring a special 
kinematical situation in which the three partons 
whose momenta are $p_{1},p_{3},p_{4}$ are collinear. In~\cite{ggam} this 
triple collinear phase space was used to compute the integral over the 
subtraction term appropriate to this kinematics, the triple 
collinear splitting function. 
The normalisation factor relating triple collinear phase space and tripole 
phase space is just $(1-y_{134})^{1-2\e}$.
It follows clearly that in the triple collinear limit where $y_{134} \to 0$, 
the tripole phase space factorisation is appropriate.

The volume of the four-parton phase space is well-known.
However, we shall now rederive it using the tripole parametrisation
described above, to demonstrate an application of the 
factorisation formula (\ref{eq:fac}) and to illustrate the 
necessary variable  transformations.

In this formula, $-\Delta'_{4}$  
can be written as a quadratic in $y_{13}$, the first 
variable to be integrated over. We obtain
$$
-\Delta'_{4}=(1-z_{1})^2(y_{13}-y_{13,a})(y_{13,b}-y_{13})\;,
$$
with $y_{13,a},y_{13,b}$ being the roots of the quadratic. 
Using
\begin{displaymath}
y_{13}=(y_{13,b}-y_{13,a})\chi +y_{13,a},
\end{displaymath}
one has
\begin{eqnarray*} 
\int (-\Delta'_{4})^{-1/2-\e}\, {\rm d}y_{13}  
&=&(1-z_1)^{-1-2\e}\, \left(y_{13,b}-y_{13,a}\right)^{-2\e}
\int_{0}^{1} {\rm d}\chi \left[\chi(1-\chi)\right]^{-1/2-\e}\;.
\end{eqnarray*}
The root difference $\left(y_{13,b}-y_{13,a}\right)$ satisfies the relation
%is related to the square root of the discriminant $\delta_{y_{13}}$ reading
\begin{displaymath}
(1-z_1)^4\,(y_{13,b}-y_{13,a})^2=16\,z_{1}\,z_{2}\,y_{14}\,(1-z_{1}-z_{2})
\left(y_{134}(1-z_1)-y_{14}\right)\equiv \delta_{y_{13}}.
\end{displaymath}
Requiring that the Gram determinant $-\Delta'_{4}$ is positive suggests
the following change of variables
\begin{eqnarray*} 
y_{14}&=&y_{134}(1-z_{1})v\;,
\\
z_{2}&=&(1-z_{1})t\; .
\end{eqnarray*}
with $t,v$ ranging between $0$ and 1.
Then  $\delta_{y_{13}}$ 
completely factorises in these new variables:
\begin{displaymath}
\delta_{y_{13}}=16\;y_{134}^{2}\;(1-z_1)^{4}\, 
z_1 \,t\, (1-t)\, v\,(1-v) \;,
\end{displaymath}
such that the volume of the four-parton phase space then becomes: 
\begin{eqnarray}
\int \d PS_{4}&=&\int \d PS_{2} \; (q^2)^{2-2\e} \;2^{-10}\;\pi^{-5+2\e}
\frac{1}{\Gamma(1-2\e)}\; \nonumber \\
& &\frac{\Gamma^2(1/2-\e)}{\Gamma(1-2\e)}
\int_{0}^{1}{\rm d}y_{134}(1-y_{134})^{1-2\e} \;y_{134}^{1-2\e}
\int_{0}^{1}{\rm d}z_{1} \;(1-z_{1})^{1-2\e}\;z_{1}^{-\e}\nonumber \\
& & \int_{0}^{1}{\rm d}t \;t^{-\e}\;(1-t)^{-\e}
\int_{0}^{1}{\rm d}v \;v^{-\e}\;(1-v)^{-\e}.
\end{eqnarray}
The integrations can be performed
straightforwardly. Finally the volume of the
four-parton phase space reads:
$$
R_4 = P_4 = \int \d PS_{4}=P_{2}\; 
 \frac{(4\pi)^{2\e}}{2^{8}\pi^4}\, 
(q^2)^{2-2\e}\,
\frac{\Gamma^3(1-\e)\Gamma(2-2\e)}{\Gamma(3-3\e)
\Gamma(4-4\e)}  \; ,  
$$
where the volume of the two-particle phase space
$P_2$ (\ref{eq:twophase}) has been factored out. 
Introducing the common normalisation factor
\begin{equation}
S_\Gamma = P_2\, \left(\frac{(4\pi)^{\e}}{16\pi^2\,\Gamma(1-\e)
}\right)^2\, , 
\end{equation}
one arrives at 
\begin{eqnarray}
R_4 &=& S_\Gamma (q^2)^{2-2\e}\, \frac{\Gamma^5(1-\e)\Gamma(2-2\e)}
{\Gamma(3-3\e)\Gamma(4-4\e)}  
\setcounter{myr4}{\value{equation}}
\label{eq:r4master}  \\
&= &
S_\Gamma\, 
(q^2)^{2-2\e}\,
\Bigg[ \frac{1}{12} + \frac{59}{72}\,\e +  \e^2  \left(
\frac{2263}{432}-\frac{\pi^2}{9} \right) + \e^3 \left(
\frac{72023}{2592} -\frac{59\pi^2}{54}-\frac{13\zeta_3}{6}\right)\nonumber \\
&& \hspace{2cm}
+\e^4 \left(\frac{2073631}{15552} -\frac{2263\pi^2}{324}
 -\frac{767\zeta_3}{36}+\frac{\pi^4}{1080}
\right)
+ {\cal O}(\e^5) \Bigg]\;.
\end{eqnarray}

\subsection{Calculation of $R_{8,a}$}

Using the factorisation of $\d PS_{4}$ into the tripole
phase space $\d PS_{T}$ and $P_{2}= \int \d PS_2$, 
the master integral $R_{8,a}$
defined by
$$
R_{8,a}= \int \d PS_{4} \frac{1}{s_{13}\;s_{14}\;s_{23}\;s_{24}} 
$$
can be written as follows:
\begin{eqnarray}
R_{8,a}&=&P_{2} \;
\int\d PS_{T} \;
(1-y_{134})^{-2}\,(q^2)^{-4}\;\frac{1}{y_{13}\,y_{14}\,z_{1}\,z_{2}} 
\nonumber \\
&=&  P_{2}\;(q^2)^{-2-2\e} \;2^{-10}\;\pi^{-5+2\e}
\frac{1}{\Gamma(1-2\e)}\; \nonumber \\ \nonumber\\
& &\int (1-y_{134})^{-1-2\e} (-\Delta'_{4})^{-1/2-\e} \Theta(-\Delta'_{4})\;
{\rm d}y_{13}\; {\rm d}y_{14}\;
{\rm d}z_{1} \; {\rm d}z_{2}\;{\rm d}y_{134} 
\;\frac{1}{y_{13}\,y_{14}\,z_{1}\,z_{2}} \;.
\end{eqnarray}

To perform the integrations, one  factorises the
 Gram determinant
 using the same change of variables as for the evaluation of $R_{4}$.
However due to the presence of the invariants in the denominator of
the integrands, the integrations are more complicated. 
Indeed, hypergeometric functions of 
involved rational arguments occur at different stages of these 
five subsequent analytic integrations.
Using several non-linear 
transformations~\cite{bateman,thesis}  on these functions where appropriate,
all integrals can be carried out in a closed analytic form, yielding 
finally
\begin{eqnarray}
R_{8,a}&=&S_\Gamma \; (q^2)^{-2-2\e} \;
 \bigg[ \frac{6}{\e^4}\frac{\Gamma^5(1-\e)\Gamma(1-2\e)}
{\Gamma(1-3\e)\Gamma(1-4\e)}
\;_4F_3\left(1,-\e,-\e,-\e;-3\e,1-\e,1+\e;1 \right)
\nonumber \\
& & \hspace{2.1cm}
-\frac{1}{\e^4}\, \frac{\Gamma^3(1-\e)\Gamma(1+\e)\Gamma^3(1-2\e)}
{\Gamma^2(1-4\e)}
\;_3F_2\left(-2\e,-2\e,-2\e;-4\e,1-2\e;1 \right)  \bigg].
\end{eqnarray}
Expanded up to finite order in $\e$, $R_{8,a}$ reads
\begin{equation}
R_{8,a}=S_\Gamma \; (q^2)^{-2-2\e} 
\; \bigg[\frac{5}{\e^4}\;-\frac{20
\pi^2}{3\e^2}-\frac{126\zeta_{3}}{\e} + \frac{7\pi^4}{18}
+ {\cal O}(\e)  \bigg].
\setcounter{myr8a}{\value{equation}}
\label{eq:r8amaster}
\end{equation}

\subsection{Unitarity Relation and Calculation of  $R_6$ and $R_{8,b}$}
\label{sec:unit}

The use of unitarity to obtain relations between loop integrals and 
phase space integrals can be found in various contexts in the literature. 
In \cite{bdkunit}, unitarity is exploited to 
compute one-loop integrals from known single 
unresolved phase space integrals.
To do the inverse, i.e.\ to derive phase space integrals from 
multi-loop integrals, has been exploited for example 
in~\cite{Bassetto:1998uv}. In~\cite{soper}, it was moreover attempted 
to implement the unitarity cancellation of infrared singularities to
compute jet observables purely numerically as appropriately weighted 
loop integrals.

As outlined in Section~\ref{sec:kin}, one can view the phase space  
integrals as particular cuts of multi-loop diagrams. In the case 
considered here, the  four-particle phase space master integrals are 
particular cuts of massless three-loop two-point functions, which were 
studied extensively in the literature~\cite{kataev,chet2,mincer}. 
In particular, all master integrals appearing 
in these three-loop two-point
functions are either known exactly or to a very high order in 
$\e$. These three master integrals are~\cite{kataev,chet2}:
\begin{eqnarray}
I_4 &=& \parbox{1.4cm}{\epsfig{file=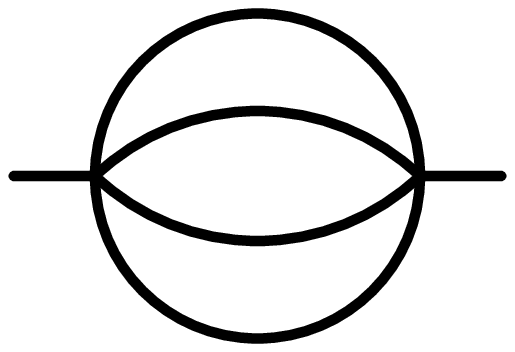,width=1.4cm}}= 
\frac{(4\pi)^{3\e}}{\left(16\pi^2\right)^3}\, 
\frac{\Gamma(1+3\e)\Gamma(1-3\e)\Gamma^4(1-\e)}{3\e\,\Gamma(3-3\e)
\Gamma(4-4\e)}\,
 (-q^2)^{2-3\e} \;, \label{eq:I4}\\
I_6 &=& \parbox{1.4cm}{\epsfig{file=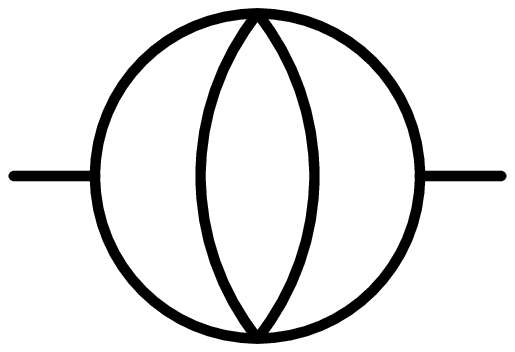,width=1.4cm}}=
\frac{(4\pi)^{3\e}}{\left(16\pi^2\right)^3}\, 
\frac{\Gamma^6(1-\e)\Gamma^3(1+\e)}
{3\e^3\,\Gamma^3(2-2\e)}\,
\Bigg[1+\e+\e^2+\e^3\left(14\zeta_3-7\right)\nonumber\\ &&\hspace{2.5cm}
+\e^4\left(-67+14\zeta_3+\frac{21\pi^4}{90} \right) + {\cal O} (\e^5) \Bigg]
 (-q^2)^{-3\e} \label{eq:I6}\\
I_8 &=& \parbox{1.4cm}{\epsfig{file=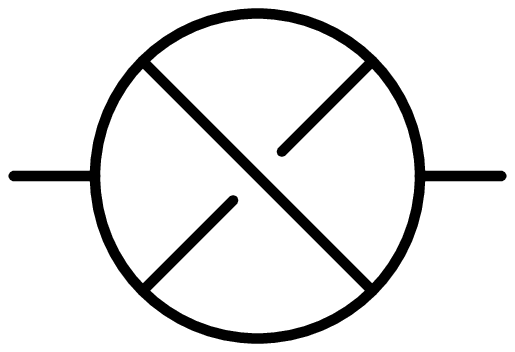,width=1.4cm}}= {\cal O}(\e^0)\;.
\end{eqnarray}

The optical theorem relates the imaginary part of a loop diagram 
to the sum over all its cuts in the form of a unitarity relation:
$$
2\; {\rm Im}\quad \parbox{1.4cm}{\epsfig{file=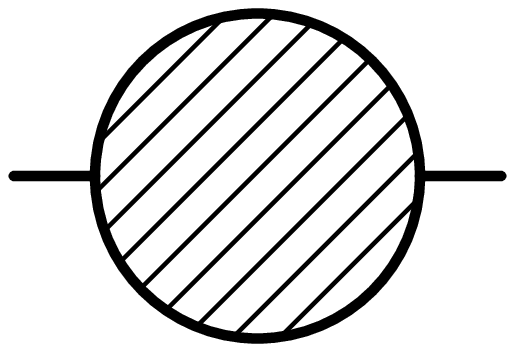,width=1.4cm}} 
=\sum_i \parbox{1.4cm}{\epsfig{file=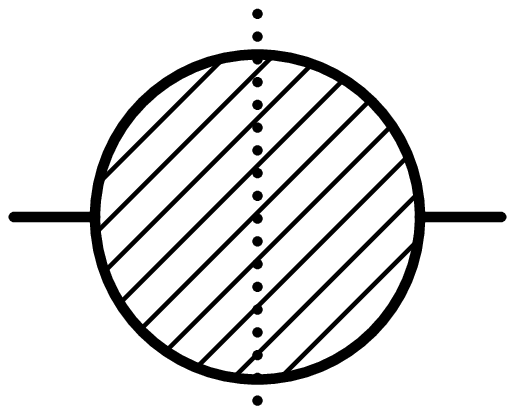,width=1.4cm}} \;,
$$
where $i$ enumerates all possible ways to cut the vacuum polarisation
diagram into two connected amplitudes under the condition that 
the final states resulting from the cut lines belong to 
physical processes. 
It has to be kept in mind that this equation relates matrix elements, not 
just loop and phase space integrals. In the case of the master integrals 
considered here, one obtains the corresponding matrix elements by 
considering a scalar theory involving 
scalar three- and four-point vertices and two-point as well as four-point 
couplings to an external scalar current. Accordingly, one 
has to include a factor $i$ for each propagator and a factor 
$-i$ for each vertex 
to obtain the desired relations among master integrals.

If all but one of the cuts are known, the 
unitarity relation can be used to 
infer the remaining unknown cut. 
In the case considered here, the three-loop two-point functions 
can have 
\begin{enumerate}
\item two-particle cuts: two-loop vertex function or product of 
two one-loop vertex functions  
integrated over two-particle 
phase space.
\item three-particle cuts: one-loop four-point function integrated over
three-particle phase space.
\item four-particle cuts: tree-level matrix element integrated over 
four-parton phase space. 
\end{enumerate}
As we shall see below, the functions appearing in the two-particle 
cuts are well known from other calculations, and the three-particle 
cuts can be computed more easily than the four-particle cuts. The 
unitarity relation following from the optical theorem
can therefore be used to determine 
the remaining four-particle phase space master integrals $R_6$ and $R_{8,b}$. 
Before computing these, we illustrate the application of the method 
on the rederivation of $R_4$. 

The unitarity relation relevant to $R_4$ reads pictorially:
$$
2\; {\rm Im}\quad \parbox{1.4cm}{\epsfig{file=I4.ps,width=1.4cm}} =\parbox{1.4cm}{\epsfig{file=R4.ps,width=1.4cm}} \;,
$$
which yields
\begin{equation}
2\, {\rm Im}\; I_4 = R_4.
\end{equation}
Using the relation
\begin{equation}
{\rm Im}\left( (-1)^{-n\e}\, \Gamma(1+n\e)\,\Gamma(1-n\e) \,
\frac{1}{n\e} \right) = \pi \,,
\end{equation}
which is valid to all orders in $\e$, one can extract $R_4$ from the 
value of $I_4$ listed in (\ref{eq:I4}):
\begin{displaymath}
R_4  =  S_\Gamma (q^2)^{2-2\e}\, \frac{\Gamma^5(1-\e)\Gamma(2-2\e)}
{\Gamma(3-3\e)\Gamma(4-4\e)} 
\end{displaymath}
in accordance with the result obtained by explicit computation in 
(\ref{eq:r4master}) above.

To determine $R_6$, we use the unitarity relation obtained for
$I_6$:
$$
2\; {\rm Im}\quad \parbox{1.4cm}{\epsfig{file=I6.ps,width=1.4cm}} = 
2\; \mbox{Re }\parbox{1.4cm}{\epsfig{file=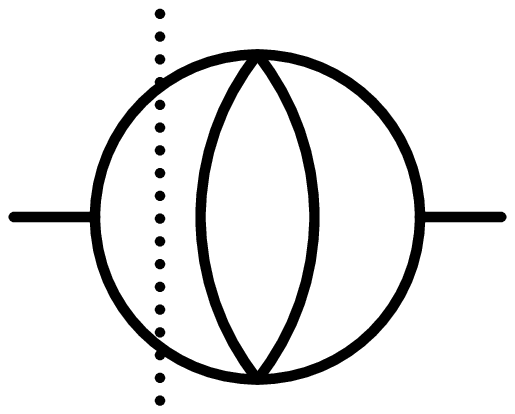,width=1.4cm}} 
+2\; \parbox{1.4cm}{\epsfig{file=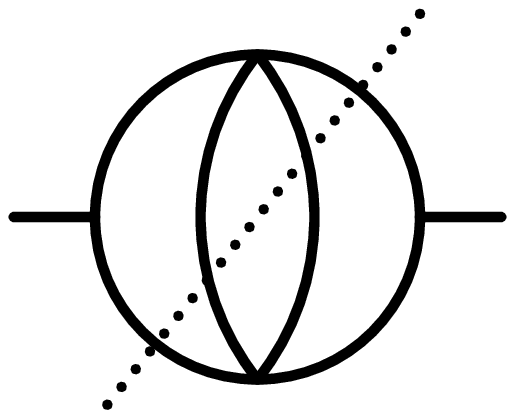,width=1.4cm}} \;,
$$
reading
\begin{equation}
2\, {\rm Im}\; I_6 = - 2\,P_2 \, {\rm Re} A_4 + 2\, R_6\;.
\label{eq:i6rel}
\end{equation}
The two-loop vertex integral $A_4$ is known from form-factor 
calculations\cite{kl}, and appeared also in the reduction of 
on-shell and off-shell massless two-loop four-point 
functions\cite{twolfourp}:
\begin{eqnarray*}
A_4 &=& 
\left(\frac{(4\pi)^{\e}}{16 \pi^2}\, \frac{ \Gamma (1+\e)
    \Gamma^2 (1-\e)}{ \Gamma (1-2\e)}
\right)^2\, \left( -q^2 \right)^{-2\e}\, 
\bigg[ -\frac{1}{2\e^2} - \frac{5}{2\e} - \left(  \frac{19}{2}
+ \frac{\pi^2}{6} \right) - \left(\frac{65}{2}
+ \frac{5\pi^2}{6} - 2\zeta_3
\right)\e \\ 
&&\hspace{6.2cm}
- \left( \frac{211}{2} + \frac{19\pi^2}{6} -10\zeta_3 \right)\e^2
+ {\cal O}(\e^3) \bigg]\;.
\end{eqnarray*}
Therefore, $R_6$ can be immediately read off from (\ref{eq:i6rel}) to be
\begin{eqnarray}
R_6 & = & 
S_\Gamma \,
(q^2)^{-2\e}\,
 \Bigg[ -1 + \frac{\pi^2}{6} + \e  \left(
-12+\frac{5\pi^2}{6}+9 \zeta_3 \right)\nonumber \\
   &&  \hspace{2cm} + \e^2  \left(-91+\frac{9\pi^2}{2}+45\zeta_3+\frac{61\pi^4}{180}\right)
+ {\cal O}(\e^3) \Bigg]\;.
\setcounter{myr6}{\value{equation}}
\label{eq:r6master}
\end{eqnarray}

In order to obtain $R_{8,b}$, we employ
\begin{eqnarray}
2\; {\rm Im}\quad \parbox{1.4cm}{\epsfig{file=I8.ps,width=1.4cm}} &=& 
2\; \mbox{Re }\parbox{1.4cm}{\epsfig{file=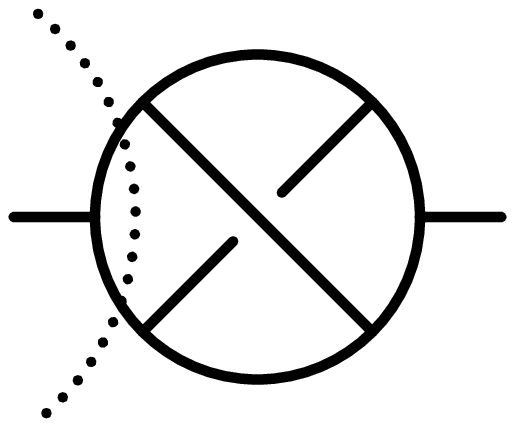,width=1.4cm}} 
+4\; \mbox{Re }\parbox{1.4cm}{\epsfig{file=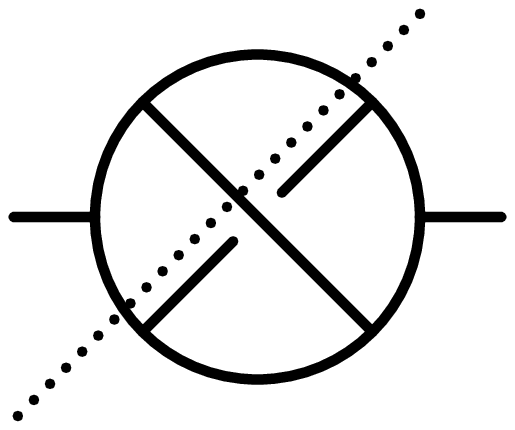,width=1.4cm}} 
+\; \parbox{1.4cm}{\epsfig{file=R8a.ps,width=1.4cm}} 
+2\; \parbox{1.4cm}{\epsfig{file=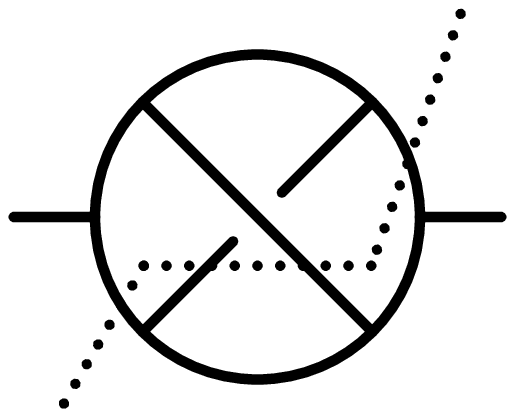,width=1.4cm}} 
+2\; \parbox{1.4cm}{\epsfig{file=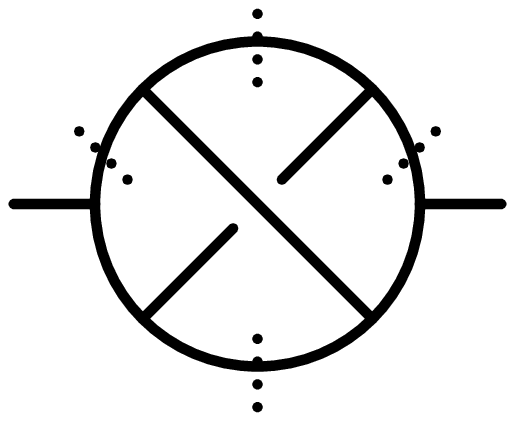,width=1.4cm}} 
\nn \\
2\; {\rm Im}\; I_8&=& - 2\,P_2\, \mbox{Re } A_6  + 4\, \mbox{Re }
V_8 +  R_{8,a} + 4 R_{8,b} \;.
\label{eq:I8rel}
\end{eqnarray}
Again, the crossed vertex master 
integral $A_6$ is well known from the literature~\cite{kl,twolfourp} to be
$$
A_6 = 
\left(\frac{(4\pi)^{\e}}{16 \pi^2}\, \frac{ \Gamma (1+\e)
    \Gamma^2 (1-\e)}{ \Gamma (1-2\e)}
\right)^2\, \left( -q^2 \right)^{-2-2\e}\, 
\left[ -\frac{1}{\e^4} +
\frac{5\pi^2}{6\e^2} + \frac{23}{\e}\zeta_3 + \frac{103\pi^4}{180}
+ {\cal O}(\e) \right]\;.
$$

The three-particle cut $V_8$ is the integral of the one-loop four-point 
functions over the three-particle phase space
\begin{equation}
V_8 = -i \int \d PS_3 \, \frac{1}{s_{12}} 
\; \mbox{Box}(s_{13},s_{13},s_{12})\;,
\end{equation}
with the one-loop box taking the well-known~\cite{kl} compact form
\begin{eqnarray}
\mbox{Box}(s_{13},s_{23},s_{12}) &=& 
\left(\frac{(4\pi)^{\e}}{16 \pi^2}\, \frac{ \Gamma (1+\e)
    \Gamma^2 (1-\e)}{ \Gamma (1-2\e)}
\right)\, \left( -q^2 \right)^{-2-\e}\,
\frac{2 i}{\e^2}\, \frac{1}{y_{13}y_{23}} \nonumber\\
&& \Bigg[\left(\frac{y_{13}y_{23}}{1-y_{13}}\right)^{-\e} 
\;_2F_1\left(-\e,-\e;1-\e;\frac{1-y_{13}-y_{23}}{1-y_{13}}\right) \nonumber \\
&&+ \left(\frac{y_{13}y_{23}}{1-y_{23}}\right)^{-\e} 
\;_2F_1\left(-\e,-\e;1-\e;\frac{1-y_{13}-y_{23}}{1-y_{23}}\right) \nonumber \\
&&-\left(\frac{y_{13}y_{23}}{(1-y_{13})(1-y_{23})}\right)^{-\e} 
\;_2F_1\left(-\e,-\e;1-\e;\frac{1-y_{13}-y_{23}}{(1-y_{13})(1-y_{23})}\right) 
 \Bigg]\; ,\nn
\end{eqnarray}
where in this case
$q^2 = s_{12}+s_{13}+s_{23}$ and $y_{ij} = s_{ij}/q^2$.
To perform the integration, the three-particle phase space is 
most conveniently parametrised in terms of the $y_{ij}$ (see Appendix):
$$
\d PS_3 = (2\pi)^{-5+4\e}\, 2^{-5+2\e}\, (q^2)^{1-2\e} \,\d \Omega_{3-2\e} 
\d \Omega_{2-2\e} \, \left(y_{12}y_{13}y_{23}\right)^{-\e} \, \d y_{12}
\, \d y_{13} \, \d y_{23} \, \delta\left(1-y_{12}-y_{13}-y_{23}\right)\;.
$$
Similar integrals were considered in the literature in the context of 
the integration of subtraction terms appropriate to single unresolved 
limits of one-loop amplitudes~\cite{kos1,wz1}. 
While the first two terms can in fact be straightforwardly 
integrated over the three-particle phase space, more care is required for the 
last term. 
We have used  several analytic continuations of hypergeometric 
functions~\cite{bateman} 
at intermediate stages where appropriate. Furthermore,   
some of the integrals could 
only be carried out on the series representation of  hypergeometric
functions, the resulting expressions
were resummed using {\tt Mathematica}~\cite{mathematica}. 
One finally arrives at
\begin{equation}
\mbox{Re }V_{8} =
S_\Gamma\,(q^2)^{-2-2\e}\,\Bigg[ -\frac{5}{2\e^4}+\frac{9\pi^2}{2\e^2}
+\frac{89\zeta_3}{\e}+\frac{13\pi^4}{180}+ {\cal O}(\e)
\Bigg]\;.
\end{equation}
As a result, one infers from (\ref{eq:I8rel}):
\begin{equation}
R_{8,b} = 
S_{\Gamma}
(q^2)^{-2-2\e}\,
\Bigg[ \frac{3}{4\e^4}-\frac{17\pi^2}{12\e^2}
-\frac{44\zeta_3}{\e}-\frac{61\pi^4}{60}+ {\cal O}(\e)\Bigg]\;.
\setcounter{myr8b}{\value{equation}}
\label{eq:r8bmaster}
\end{equation}

Finally, we can also employ the unitarity relation to check 
the correctness of the reduction developed in Section~\ref{sec:red} by 
computing a reducible integral directly through the unitarity relation and 
comparing it with the result obtained from the reduction procedure. 
For this exercise, we choose
\begin{equation}
R_{8,r} \equiv \parbox{1.4cm}{\epsfig{file=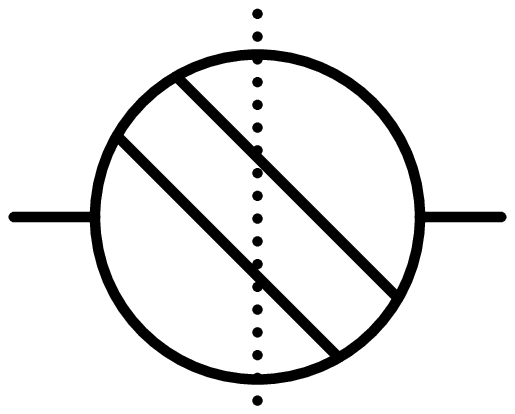,width=1.4cm}}
= \int \d PS_4 \frac{1}{s_{13}s_{134}s_{24}s_{234}}\;,
\end{equation}
which reduces to
\begin{eqnarray}
R_{8,r} &=& 
\frac{(1-7\e+18\e^2)\,(2-3\e)\,(3-4\e)}{2\e^4}\; R_4
+ \frac{3\,(1-2\e)\,(1-3\e)}{2\e^2} \; R_6\; \nn \\
&=& S_\Gamma\; 
(q^2)^{-2-2\e}\;\Bigg[ \frac{1}{4\e^4}-\frac{\pi^2}{12\e^2}
+\frac{7\zeta_3}{\e}+\frac{23\pi^4}{45}+ {\cal O}(\e)\Bigg]\;.
\label{eq:r8red}
\end{eqnarray}
This integral is related to the imaginary part of\cite{kataev,chet2} 
\begin{equation}
I_{8,r} = \parbox{1.4cm}{\epsfig{file=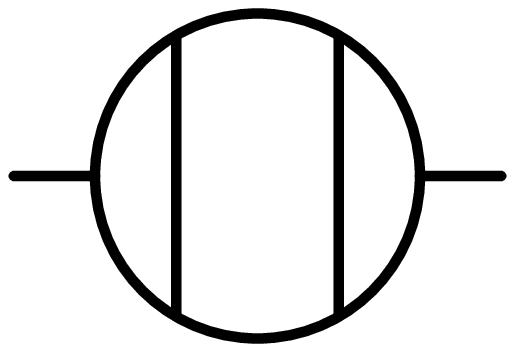,width=1.4cm}}= {\cal O}(\e^0)\;,
\end{equation}
for which the unitarity relation reads 
\begin{eqnarray}
2\; {\rm Im}\quad \parbox{1.4cm}{\epsfig{file=I8r.ps,width=1.4cm}} &=& 
2\; \mbox{Re }\parbox{1.4cm}{\epsfig{file=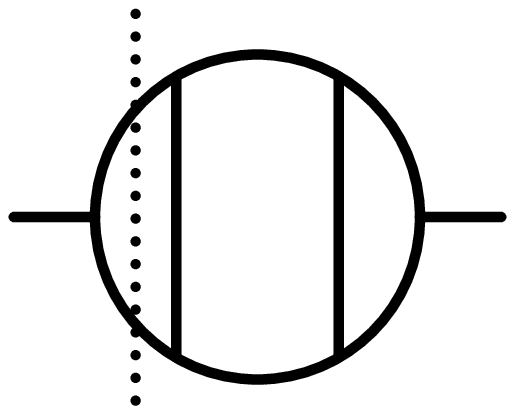,width=1.4cm}} 
+4\; \mbox{Re }\parbox{1.4cm}{\epsfig{file=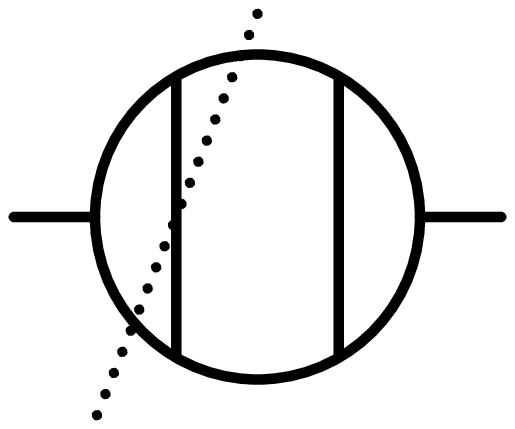,width=1.4cm}} 
+\; \parbox{1.4cm}{\epsfig{file=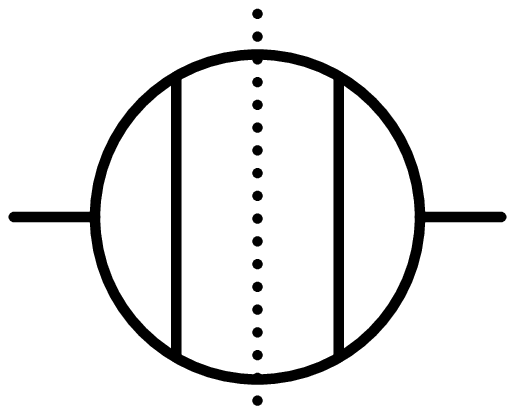,width=1.4cm}} 
+2\; \parbox{1.4cm}{\epsfig{file=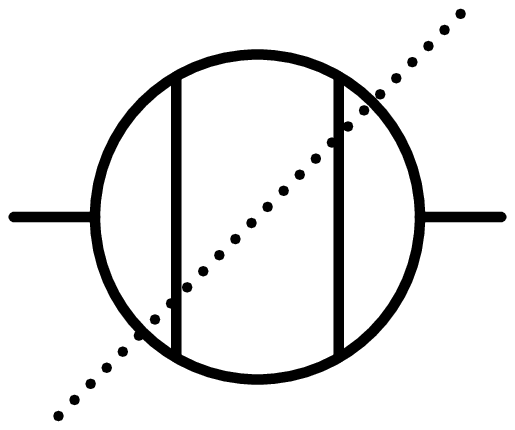,width=1.4cm}} 
\nn \\
2\; {\rm Im}\; I_{8,r}&=& - 2\,P_2\, \mbox{Re } A_{6,r}  + 4\, \mbox{Re }
V_{8,r} +  |A_{3,r}|^2 \,P_2  + 2 R_{8,r} \;.
\label{eq:I8red}
\end{eqnarray}
The terms in this relation are easily evaluated as 
\begin{eqnarray*}
A_{6,r} &=& \left(\frac{(4\pi)^{\e}}{16 \pi^2}\, \frac{ \Gamma (1+\e)
    \Gamma^2 (1-\e)}{ \Gamma (1-2\e)}
\right)^2\, \left( -q^2 \right)^{-2-2\e}
\left[ -\frac{1}{4\e^4} -
\frac{\pi^2}{4\e^2} - \frac{6}{\e}\zeta_3 - \frac{3\pi^4}{20}
+ {\cal O}(\e) \right]\;,\\
\mbox{Re }V_{8,r} &=& 
S_\Gamma \,(q^2)^{-2-2\e}\,\Bigg[ -\frac{1}{2\e^4}+\frac{\pi^2}{6\e^2}
-\frac{5\zeta_3}{\e}-\frac{5\pi^4}{36}+ {\cal O}(\e)\Bigg]\;,
\\
|A_{3,r}|^2&=& \left(\frac{(4\pi)^{\e}}{16 \pi^2}\, \frac{ \Gamma (1+\e)
    \Gamma^2 (1-\e)}{ \Gamma (1-2\e)}
\right)^2\, \left( q^2 \right)^{-2-2\e}
 \frac{1}{\e^4} \;,
\end{eqnarray*}
such that together with (\ref{eq:r8red}), the unitarity relation 
(\ref{eq:I8red}) is fulfilled identically. Since only the evaluation 
of $R_{8,r}$ relied on the reduction of phase space integrals, this provides 
a strong check on the correct implementation of the reduction algorithm 
described in Section~\ref{sec:red}.
%%%%%%%%%%%%%%%%%%%%%%%%%%%%%%%%%%%%%%%%%%%%%%%%%%%%%%%%%%%%%%%%

\section{Numerical Calculation of Master Integrals}
\setcounter{equation}{0}
\label{sec:num}
As outlined in~\cite{Heinrich:2002rc}, 
the method of sector decomposition~\cite{hepp} 
to isolate the (infrared) divergences 
from a given parameter integral and to calculate the resulting 
pole coefficients numerically can be applied not only to 
multi-loop integrals~\cite{Binoth:2000ps,Binoth:2003ak}, 
but also to phase space integrals. 
To this aim, the phase space for the $1\to 4$ particle reaction 
considered here has to be cast into a convenient parameter 
integral form, as will be explained below.  

The starting point is the expression (\ref{eq:r4}) for the four-particle 
phase space. This form, where the integration variables are the kinematic 
invariants $s_{ij}$, is most convenient  
for the sector decomposition algorithm since it is maximally symmetric 
in the integration variables.  
In addition, the potential infrared divergences 
stem a priori only from the limits $s_{ij}\to 0$ 
of a certain set of invariants  
(appearing in the denominator of some matrix element), 
%upon integration over the phase space)
that is, they are located at the origin of parameter space.

Further, we rescale the Mandelstam invariants, in order to deal with
dimensionless parameters, by the definitions
\begin{eqnarray*}
&&x_1=s_{12}/q^2, x_2=s_{13}/q^2, 
x_3=s_{23}/q^2,\\
&&x_4=s_{14}/q^2,x_5=s_{24}/q^2,
x_6=s_{34}/q^2
\end{eqnarray*}
to arrive at
\begin{eqnarray}
\int \d PS_4&=&C_{\Gamma}
\,(q^2)^{\frac{3d}{2}-4}
\int\Big\{\prod\limits_{j=1}^6 \d x_j\,\Theta(x_j)\Big\}\,\delta(1-\sum\limits_{i=1}^6
x_i)\,
\Big[-\lambda(x_1x_6,x_2x_5,x_3x_4)\Big]^{\frac{d-5}{2}}
\Theta(-\lambda)\;,\label{fi4y}
\end{eqnarray}
where 
$$
C_{\Gamma}=(2\pi)^{4-3d}V(d-1)V(d-2)V(d-3)\,2^{1-2d}\;,$$
$\lambda(x_1x_6,x_2x_5,x_3x_4)$ is the K\"allen function already introduced in
(\ref{eq:delta4}) and $V(d)$ is
the $d$-dimensional volume defined in 
(\ref{eq:omega}). 
%$\lambda_x(x_1x_6,x_2x_5,x_3x_4)=
%\lambda(s_{12}s_{34},s_{13}s_{24},s_{14}s_{23})/(Q^2)^4$.

After these transformations, we obtain for the integrals $R_6$ and 
$R_{8a,b}$ 
\begin{eqnarray}
R_6&=&(q^2)^{-2}\int \d PS_4\,\frac{1}{(x_2+x_4+x_6)(x_3+x_5+x_6)}\;,\\
R_{8,a}&=&(q^2)^{-4}\int \d PS_4\,\frac{1}{x_2x_3x_4x_5}\;,\\
R_{8,b}&=&(q^2)^{-4}\int \d PS_4\,\frac{1}{x_2x_3(x_2+x_4+x_6)(x_3+x_5+x_6)}\;.
\end{eqnarray}
The variables in (\ref{fi4y}) are constrained not only by the 
momentum conserving delta-distribution $\delta(1-\sum\limits_{i=1}^6 x_i)$, 
but also by the  requirement 
\begin{equation}
\lambda(x_1x_6,x_2x_5,x_3x_4)\le 0\Leftrightarrow x_1^2x_6^2+x_2^2x_5^2+x_3^2x_4^2-
2\,(x_1x_3x_4x_6+x_2x_3x_4x_5+x_1x_2x_5x_6)\leq 0\;.\label{gram}
\end{equation}
Therefore the integration variables are hard to factorise and 
disentangling the overlapping pole structure in general is a highly nontrivial 
task. That is where the iterated decomposition into sectors 
until a non-overlapping form is reached shows its virtues. 

\subsection{Sector decomposition}
First we decompose the original integral into a sum of six 
integrals corresponding to the six "primary sectors" created by inserting unity 
in the form 
$$
1=
\sum_{j=1}^6\prod\limits_{\stackrel{k=1}{k\not=j}}^6 \Theta(x_j-x_k)\;.
$$
In each  primary sector $j$ created in this way 
we transform variables according to
$$
x_k=\left\{
\begin{array}{ll}
x_j\,t_k&\mbox{if } k\not=j\\
x_j&\mbox{if } k=j
\end{array}
\right.
$$
and integrate out $x_j$ using the 
momentum conserving $\delta$-distribution,  
to arrive 
at the following form for the phase space integral  
(where we choose $j=1$ to give a concrete example):
\begin{eqnarray}
\int \d PS_4^{(sec1)}&=&
C_{\Gamma}\,(q^2)^{\frac{3d}{2}-4}
\Big\{\prod\limits_{i=2}^6 \int \d t_i\,\Theta(t_i)\,\Theta(1-t_i)\Big\}\,
A_{(sec1)}^{4-2d}\nn\\
&&\Big[-\lambda^{(sec1)}(t_6,t_2t_5,t_3t_4)\Big]^{\frac{d-5}{2}}
\,\Theta(-\lambda^{(sec1)})\;,\\
A_{(sec1)}&=&1+\sum\limits_{i=2}^{6} t_i\;,\nn\\
-\lambda^{(sec1)}(t_6,t_2t_5,t_3t_4)&=&t_6^2+t_2^2t_5^2+t_3^2t_4^2-
2\,(t_3t_4t_6+t_2t_3t_4t_5+t_2t_5t_6)\;.\label{d41}
\end{eqnarray}
Solving the equation $-\lambda^{(sec1)}=0$ for $t_6$ 
(or, in general, in sector $j$, for $t_{7-j}$,
which occurs simply as $t_{7-j}^2$
 in  $\lambda^{(secj)})$\,: 
% in eq.~(\ref{d41})):
$$
t_6^{\pm}=t_2t_5+t_3t_4\pm 2\sqrt{t_2t_3t_4t_5}=
(\sqrt{t_2t_5}\pm\sqrt{t_3t_4})^2
$$
and substituting 
\begin{equation}
t_6=\tilde t_6\,(t_6^+-t_6^-)+t_6^-
=\tilde{t}_6\,4\sqrt{t_2t_3t_4t_5}+t_2t_5+t_3t_4- 2\sqrt{t_2t_3t_4t_5}\;,
\label{t6}
\end{equation}
where now
$$
-\lambda^{(sec1)}=\tilde{t}_6(1-\tilde{t}_6)(t_6^+-t_6^-)^2=
16\,t_2t_3t_4t_5\,\tilde{t}_6(1-\tilde{t}_6)\;,
$$
leads to
\begin{eqnarray}
\int \d PS_4^{(sec1)}&=&
S_{\Gamma}\,(q^2)^{d-2}\,\frac{\Gamma(1-2\e)}{\Gamma^2(\frac{1}{2}-\e)}
\Big\{\prod\limits_{i=2}^5\int \d t_i\,\Theta(t_i)\,\Theta(1-t_i)\Big\}\,
[t_2t_3t_4t_5]^{-\e}\nn\\
&& \int \d\tilde{t}_6\,
[\tilde{t}_6(1-\tilde{t}_6)]^{-\frac{1}{2}-\epsilon}
\,\Theta(\tilde{t}_6)\,\Theta(1-[\tilde
t_6\,(t_6^+-t_6^-)+t_6^-])\;A_{(sec1)}^{-4+4\epsilon}\;.
\label{t6tilde}
\end{eqnarray}
However, as a consequence of the transformation (\ref{t6}),   
singularities at $t_i=1$ can also occur now. For example, 
a factor $1/t_6$ that might be present in the matrix element can 
become singular for $\tilde{t}_6\to 0, t_{i=2,3,4,5}\to 1$. 
As the subtractions performed later rely on the fact that 
only the limits $t_i\to 0$ can lead to a singula\-rity, 
these potential singularities  are remapped to the origin in the following way:
The program checks if a (non-empty) set ${\cal T}$ exists 
such that a denominator 
vanishes if all $t_i \in {\cal T}$ are one and $\tilde{t}_k\to 0$. 
For the $t_i\in {\cal T}$, it is checked if in addition $t_i\to 0$ 
is singular. 
If yes, the integration region is split accor\-ding to 
$\int_0^1 dt_i=\int_0^{\frac{1}{2}} dt_i+\int_{\frac{1}{2}}^1 dt_i$ 
and the resulting integrals are remapped to integrals from zero to one. 
Otherwise, 
simply the transformations $t_i\to 1-t_i$ are performed for the 
$t_i\in {\cal T}$. 
Note that the limits $\{\tilde{t}_k\to 0,t_i\to 1 \,(t_i\in {\cal T})\}$ 
do in most cases {\em not} 
lead to a real $1/\e$ singularity upon integration, even if they make 
the denominator vanish, but they will be remapped nevertheless 
for the sake of numerical stability. 

These transformations as well as the sector decomposition are iterated 
if necessary, that is, if a (composite\footnote{By "composite" we mean that 
the denominator is a nontrivial function of parameters $t_i$, 
as for example $(t_2+t_4+t_6)$ present in $R_{8,b}$, and not only 
a trivial factor. Note that after the transformation (\ref{t6})
a previously trivial factor like $1/t_6$ becomes composite.}) 
denominator of the integrand still can vanish at a certain phase space point.  

The iterated sector decomposition proceeds as follows:
The program determines the minimal set 
${\cal S}=\{t_{\alpha_1},\dots ,t_{\alpha_r}\}$ 
of parameters which in the limit 
$t_{\alpha}\to 0$ make a composite denominator of the matrix 
element vanish. (The parameters $\tilde{t}$ have been renamed 
$t$).
Then the corresponding integrals are decomposed into $r$  subsectors 
according to
\begin{eqnarray}
\prod\limits_{i=1}^r \Theta(1- t_{\alpha_i})=
\sum\limits_{k=1}^r \prod\limits_{\stackrel{i=1}{i\ne k}}^r 
\Theta(t_{\alpha_k}- t_{\alpha_i})\;.
\end{eqnarray}
In each new subsector, the variables are remapped to the unit cube 
by the substitution
 \begin{eqnarray}
t_{\alpha_i} \rightarrow 
\left\{ \begin{array}{lll} 
t_{\alpha_k} t_{\alpha_i} &\mbox{for}&i\not =k \\
t_{\alpha_k}              &\mbox{for}& i=k  \end{array}\right.\;.
\end{eqnarray}
By construction,
$t_{\alpha_k}$ factorises from at least one of the denominator functions. 
The sector decomposition procedure is iterated until 
the denominators are of the form $D=1+f(t_{\alpha})$, where $f(t_{\alpha})$ is a 
non-negative function of the parameters $t_{\alpha}$. 
%These iterations produce a certain number of subsectors in each primary sector.  
Then subtractions of the singularities, which are now 
all factorised explicitly, are carried out, using 
identities like  
\begin{eqnarray*}
\int_0^1 \d t_\alpha\,t_\alpha^{-1-\kappa\e}\,{\cal F}(t_\alpha,t_{\beta\not=\alpha})=
-\frac{1}{\kappa\e}\,\,{\cal F}(0,t_{\beta\not=\alpha})
+\int_0^1 \d t_\alpha\,t_\alpha^{-1-\kappa\e}\,\bigg\{{\cal F}(t_\alpha,t_{\beta\not=\alpha})-
{\cal F}(0,t_{\beta\not=\alpha})\bigg\}\;,
\end{eqnarray*}
where $\lim_{t_{\alpha}\to 0}{\cal F}(t_{\alpha},t_{\beta\neq \alpha})$
is finite 
by construction.

After having carried out the subtractions for each $t_\alpha$, the expansion in 
$\e$ can be performed.   
The result is a Laurent series in $\e$ where the pole coefficients are 
finite functions of the parameters $t_{\alpha}$ which can be integrated numerically. 

Note that the coefficient functions 
of a certain pole,  produced by the iterated 
decomposition of a certain primary sector into subsectors, 
are all summed over 
{\em before} numerical integration in order to avoid large 
numerical errors due to large cancellations. In contrast, the 
contributions from the {\it primary} sectors (i.e. six contributions 
in the $1\to 4$ case)
are summed over only {\it after} the numerical integration. 
In this way, one has a strong check at hand if for topological 
reasons (symmetry of the considered integral) some primary sectors  
have to yield identical results, while the numerical error 
is not substantially increased.  

We would like to emphasize that the  functions occurring in the phase space 
integrals are quite different from the functions present in any 
virtual multi-loop integral after Feynman parametrisation: 
%Apart from being products of several functions, their main feature is that
They are not polynomial anymore, as they unavoidably contain square-roots 
(and fake singularities away from the origin, as discussed above). 
This is of course  not a principal problem for the algorithm, 
but has to be dealt with appropriately. 

The $\Theta$-function and the $A_{(secj)}$ term in (\ref{t6tilde}) 
become quite involved expressions after several iterations, 
but do not present any problem for the numerical integration. 

The program consists of 3 building blocks: 
The first one serves to perform the iterated sector decomposition 
until a form of the integrand is reached where all singularities 
are factored out explicitly as described above. 
The corresponding algorithm has been implemented in 
{\tt Mathematica}~\cite{mathematica}.

Then the subtractions and the expansion in $\e$ are carried out 
to obtain a set of finite functions for each pole coefficient. 
These functions are then translated into {\tt FORTRAN} functions. 
Up to this stage, the method is {\em not} "numerical" yet. 
Only the complicated structure and the large number of these functions
hampers their analytical integration. 

The third part consists in the numerical integration of these 
functions, where the Monte Carlo program {\tt BASES}~\cite{bases} is used. 
All {\tt FORTRAN} files are generated automatically using 
{\tt Maple}~\cite{maple} and {\tt perl}. 
{\tt Perl} routines also serve to automate compilation, job submission 
and summation of the results from the different primary sectors, 
such that the only "manual" input is the matrix element and 
the desired expansion level in $\e$.

\subsection{Numerical Results}

The program produces an overall number of 
13 subsectors for the integral $R_6$, 690 subsectors for $R_{8,a}$ and 
321 subsectors for $R_{8,b}$.
The integration time ranges from about 10 minutes for $R_6$ to about 24 hours 
for $R_{8,a}$ if a numerical accuracy better than 1\% is demanded. 
%Demanding less accuracy and further parallelisation speeds up the 
%calculation considerably. 

The following numerical results are obtained for the master integrals:
\begin{eqnarray}
R_4&=&S_{\Gamma}(q^2)^{2-2\e}\left[0.08335+0.81946\,\e+4.1413\e^2+
14.398\,\e^3+38.880\,\e^4+{\cal O}(\e^5)\right]\;,\\
R_6&=&S_{\Gamma}(q^2)^{-2\e}\left[0.64498+7.0423\e+40.507\e^2
+{\cal O}(\e^3)\right]\;,\\
R_{8,a}&=&S_{\Gamma}(q^2)^{-2-2\e}\left[\frac{5.0003}{\e^4}-
\frac{0.0013}{\e^3}-\frac{65.832}{\e^2}-
\frac{151.53}{\e}+ 37.552+{\cal O}(\e)\right]\;,\\
R_{8,b}&=&S_{\Gamma}(q^2)^{-2-2\e}\left[\frac{0.74986}{\e^4}-\frac{0.00009}{\e^3}-
\frac{14.001}{\e^2}-\frac{52.911}{\e}-99.031 +{\cal O}(\e)\right]\;.
\end{eqnarray}
The reducible integral $R_{8,r}$ also has been calculated numerically 
to complete the cross-checks:
\begin{equation}
R_{8,r}=S_{\Gamma}(q^2)^{-2-2\e}\left[\frac{0.24999}{\e^4}-\frac{0.00025}{\e^3}-
\frac{0.82251}{\e^2}+\frac{8.4106}{\e}+49.748 +{\cal O}(\e)\right]\;.
\end{equation}
The agreement of the numerical and analytical results is better than 1\%.

\section{Summary and Conclusions}
\setcounter{equation}{0}
\label{sec:conc}

In this paper, we studied the inclusive 
four-particle phase space integrals of ($1\to 4$)-parton QCD matrix 
elements. These integrals contain infrared divergences due to 
the soft or collinear emission of up to two partons, which occur also 
in NNLO calculations of jet observables. We demonstrated that 
these phase space integrals can be expressed as a linear combination 
of only four master integrals. 
These four master integrals were computed both
analytically and numerically up to their fourth order terms in 
dimensional regularisation, as required for NNLO calculations.
The analytic results for them are: 
\renewcommand{\theequation}{\mbox{\ref{sec:an}.\arabic{equation}}}
\begin{eqnarray}
R_4 & = & S_\Gamma\; (q^2)^{2-2\e}\, 
\frac{\Gamma^5(1-\e)\Gamma(2-2\e)}{\Gamma(3-3\e)
\Gamma(4-4\e)} \;, 
\setcounter{equation}{\value{myr4}}
\\
R_6 & = & 
S_\Gamma \;
(q^2)^{-2\e}\,
\Bigg[ -1 + \frac{\pi^2}{6} + \e  \left(
-12+\frac{5\pi^2}{6}+9 \zeta_3 \right)\nonumber \\
   &&  \hspace{2cm} + \e^2  \left(-91+\frac{9\pi^2}{2}+45\zeta_3+\frac{61\pi^4}{180}\right)
+ {\cal O}(\e^3) \Bigg]\;,
\setcounter{equation}{\value{myr6}}
\\
R_{8,a} & = & 
S_\Gamma \;
(q^2)^{-2-2\e}\,
\Bigg[\frac{5}{\e^4}-\frac{20\pi^2}{3\e^2}
-\frac{126\zeta_3}{\e}+\frac{7\pi^4}{18} + {\cal O}(\e) \Bigg]
\;,
\setcounter{equation}{\value{myr8a}}
\\
R_{8,b} & = & 
S_\Gamma \;
(q^2)^{-2-2\e}\,
\Bigg[ \frac{3}{4\e^4}-\frac{17\pi^2}{12\e^2}
-\frac{44\zeta_3}{\e}-\frac{61\pi^4}{60}+ {\cal O}(\e)\Bigg]
\;.\setcounter{equation}{\value{myr8b}}
\end{eqnarray}

The inclusive four-particle phase space integrals of QCD $1\to 4$
matrix elements may in future work be used for the construction of 
infrared subtraction terms for the real contributions to 
NNLO jet calculations. 
In the case of two-jet production in $e^+e^-$ collisions,
the $1\to 4$ matrix elements can serve immediately as subtraction
terms, since the only requirements on subtraction terms are that 
they coincide with the full matrix element in all 
infrared singular limits (which is clearly the case) and that they can
be integrated analytically over the inclusive phase space (which we
have demonstrated here). 
Constructing NNLO subtraction terms for 
the double real emission contributions to three-jet  production 
in $e^+e^-$ collisions 
from $1\to 4$ matrix elements, which can serve to locally approximate 
the full $1\to 5$ matrix element in certain limits, 
is a yet outstanding task.
To the same aim, 
one could also look for a way to relate already existing 
NNLO subtraction terms~\cite{twot,kosower,wz2}
to $1\to 4$ QCD matrix elements. 
In  this context, it should be pointed out that the frequently used 
NLO dipole subtraction terms~\cite{cs}
for multi-jet production in $e^+e^-$ annihilation can
in fact be recombined to yield $1\to 3$ QCD matrix elements. 

In the course of the calculations
presented in this paper, new analytical and numerical
methods for handling infrared divergent phase space integrals were
developed. In particular, the tripole parametrisation of the four-particle 
phase space incorporates explicitly the phase space
factorisation from an ($n$+2)-parton phase space to an $n$-parton 
phase space and a factor which accounts for the singular emission, 
as required for $n$-jet production at NNLO. This tripole 
formulation has the advantage that it avoids the introduction of  
non-linear transformations of momenta, which is crucial for the 
analytic integration. For the construction of an NNLO Monte
Carlo parton-level event generator, 
a  smooth mapping of all potentially singular regions of phase space
is essential for numerical stability. 
Such a mapping can for example be achieved using non-linear momentum 
transformations~\cite{kosower,wz2}.
However, the phase space parametrisations used to 
integrate the subtraction terms analytically and those used in the 
Monte Carlo program do not need to be the same.  

Formulating the unitarity relations between phase
space and loop integrals in a form which connects 
different master integrals enabled us to deduce three  
of the four master integrals from known multi-loop integrals, 
such that only one of them had to be calculated explicitly. 
Thus we were able to use multi-loop techniques 
in order to obtain results for complicated phase space integrals.

The iterated sector decomposition, a method 
originally developed for infrared divergent loop integrals, 
where the poles are isolated automatically and their coefficients 
are calculated numerically, was extended here to phase space integrals. 
This technique is particularly powerful 
for  ($1 \to n$)-particle reactions, where the 
phase space integrals depend only on one overall scale. 
As a consequence, 
the pole coefficients are just numbers which can be calculated
numerically once and for all. 
This method can also be useful in cases where an analytic
calculation of the phase space integrals cannot be achieved anymore, 
not only for inclusive cross sections, but also 
in view of the construction of an NNLO Monte Carlo program,
where experimental cuts  acting on the phase space boundaries
can complicate the analytic integrations.

The excellent agreement between analytical and numerical results for
the four master integrals evaluated here provides a strong 
check on both calculational approaches.

\section*{Acknowledgements}
AG and TG wish to thank Rolf Mertig for numerous instructive
discussions on 
hypergeometric functions and their usage within the computer algebra 
system {\tt Mathematica}. AG acknowledges the kind hospitality of
 the Institut  f\"ur Theoretische Physik of RWTH Aachen, 
where part of this work was performed.
GH would like to thank the University of Z\"urich for its 
kind hospitality. 
This work was supported by the Swiss National Funds 
(SNF) under contract 200021-101874.

\begin{appendix}
\renewcommand{\theequation}{\mbox{\Alph{section}.\arabic{equation}}}
\section{Massless Multi-parton Phase Space}
\setcounter{equation}{0}
\label{app:ps}

The $n$-particle phase space in the  ($1\to n$)-particle decay reaction 
kinematics ($q\to p_1 + \ldots + p_n$) reads in dimensional 
regularisation with $d=4-2\e$ space-time dimensions
\begin{equation}
\d PS_n = \frac{\d^{d-1} p_1}{2E_1 (2\pi)^{d-1}}\; \ldots \;
\frac{\d^{d-1} p_n}{2E_n (2\pi)^{d-1}}\; (2\pi)^{d} \;
\delta^d (q - p_1 - \ldots - p_n) \,.
\end{equation}
In the present work, we are in particular concerned about the two-, three-
and four-parton phase space.

In practical applications, it turns out that 
the above expression for the multi-particle phase space in terms of momenta 
is not convenient. We therefore use expressions in terms of 
kinematic invariants $s_{ij} = 2\, p_i\cdot p_j$ and 
angular volumes instead. Further, we factorise out overall rotations
of the coordinate frame in $d$ dimensions into the $d$-dimensional hypersphere
$\d \Omega_d$ with volume
\begin{equation}
V({d}) \equiv \int {\rm d}\Omega_{d}\;=\;\frac{2\pi^{d/2}}{\Gamma(d/2)}.
\label{eq:omega}
\end{equation}
The two-particle phase space in these variables simply reads 
\begin{equation}
\d PS_2 = (2\pi)^{2-d} (s_{12})^{\frac{d-4}{2}}\;
\frac{{\rm d}\Omega_{d-1}}{2^{d-1}}\;{\rm d}s_{12}\;\delta(q^2-s_{12})\;.
\end{equation}
For the three-particle phase space, one finds
\begin{equation}
\d PS_3 = (2\pi)^{3-2d}\, 2^{-1-d}\, (q^2)^{\frac{2-d}{2}} \,\d \Omega_{d-1} 
\d \Omega_{d-2} \, \left(s_{12}s_{13}s_{23}\right)^{\frac{d-4}{2}} \, \d s_{12}
\, \d s_{13} \, \d s_{23} \, \delta\left(q^2-s_{12}-s_{13}-s_{23}\right)\;.
\end{equation}
Finally, the four-particle phase space is
\begin{eqnarray}
\d PS_4 &=& 
(2\pi)^{4-3d} (q^2)^{3-\frac{d}{2}} 2^{1-2d}
(-\Delta_{4})^{\frac{d-5}{2}}\;\Theta(-\Delta_{4})
\;\delta(q^2 -s_{12}- s_{13}-s_{14}-s_{23}-s_{24}-s_{34})
%(2\pi)^{4-3d}  \frac{{(-\Delta_{4})}^{-1/2}}{q^{2}\;2^{9}}
%{\rm d}\Omega_{d-1}\;{\rm d}\Omega_{d-2}\;{\rm d}\Omega_{d-3}
%\;\delta(q^2 -s_{12}- s_{13}-s_{14}-s_{23}-s_{24}-s_{34})
\nonumber\\
& & {\rm d}\Omega_{d-1}\;{\rm d}\Omega_{d-2}\;{\rm d}\Omega_{d-3}
{\rm d}s_{12}\;{\rm d}s_{13}{\rm d}s_{14}{\rm d}s_{23}{\rm d}s_{24}
{\rm d}s_{34},
\label{eq:r4} 
\end{eqnarray}
where the Gram determinant $\Delta_{4}$ is given by
\begin{eqnarray}
+\Delta_{4} & = & 
\lambda(s_{12}s_{34},s_{13}s_{24},s_{14}s_{23})\;,\;
\lambda(x,y,z)=x^2+y^2+z^2-2(xy+xz+yz)\;.
%\bigg[{s_{12}}^{2}{s_{34}}^{2}\; +\;
%{s_{13}}^{2}{s_{24}}^{2}\;+\;{s_{14}}^{2}{s_{23}}^{2}
%\nonumber\\
%& & -2\bigg(s_{12}s_{23}s_{34}s_{14}\;+\;s_{13}s_{23}s_{24}s_{14}\;
%+\;s_{12}s_{24}s_{34}s_{13} \bigg)\bigg].
\label{eq:d4}
\end{eqnarray}

In several parts of this paper, we 
also need the total $n$-particle phase space volume
\begin{equation}
P_n = \int \d PS_n.
\end{equation}
Using the unitarity relation of Section \ref{sec:unit}, we find a 
compact expression for arbitrary $n$:
\begin{equation}
P_n = 2^{5-4n-2\e + 2n\e}\, \pi^{3-2n-\e+n\e}\, 
\frac{\Gamma^n(1-\e)}{\Gamma\left( (n-1) (1-\e) \right) 
\Gamma\left( n(1-\e) \right)} \, (q^2)^{n-2+\e-n\e}\;,
\end{equation}
in particular:
\begin{eqnarray}
P_2 & = & 2^{-3+2\e}\, \pi^{-1+\e}\, \frac{\Gamma(1-\e)}{\Gamma(2-2\e)}\,
 (q^2)^{-\e} \; ,
\label{eq:twophase}\\
P_3 & = & 2^{-7+4\e}\, \pi^{-3+2\e}\, \frac{\Gamma^3(1-\e)}{\Gamma(2-2\e)
\Gamma(3-3\e)}\,
 (q^2)^{1-2\e} \; ,\\
P_4 & = & 2^{-11+6\e}\, \pi^{-5+3\e}\, \frac{\Gamma^4(1-\e)}{\Gamma(3-3\e)
\Gamma(4-4\e)}\,
 (q^2)^{2-3\e} \;.
\end{eqnarray}

\end{appendix}

\end{document}